\documentclass[runningheads]{llncs}
\usepackage[T1]{fontenc}
\usepackage{graphicx}
\usepackage{xcolor,colortbl}
\usepackage[ruled,vlined]{algorithm2e}
\SetKwInput{KwInput}{Input}
\SetKwInput{KwOutputput}{Output}
\usepackage[most]{tcolorbox}
\usepackage{tabularx}
\usepackage{newfloat}
\usepackage{listings}
\usepackage{soul}
\usepackage{color}
\usepackage{caption} 
\usepackage{multirow}

\newcommand{\AlgoName}{\textsc{LocalIntel}\xspace}

\newcommand{\glb}{\mathcal{G}\xspace}
\newcommand{\loc}{\mathcal{L}\xspace}
\newcommand{\completion}{\mathcal{C}\xspace}
\newcommand{\query}{\mathcal{Q}\xspace}

\newtcolorbox{boxEnv}{
    center,
    left=0 mm,
    top = 0.25 mm,
    right = 0mm,
    bottom =0.25 mm,
    colframe=gray!90!black,
    colback=black!5!white, 
    boxrule=0.5pt,
    title = Problem Statement,
}

\newcommand{\SectionIntroduction}{{Introduction}}
\newcommand{\SectionObjective}{{Research Objective \& Theoretical Foundations}}
\newcommand{\SectionSystemDesc}{{Framework}}
\newcommand{\SectionExperiment}{{Experiment and Evaluation}}
\newcommand{\SectionBackground}{{Related Works}}

\begin{document}
\title{LocalIntel: Generating Organizational Threat Intelligence from Global and Local Cyber Knowledge}
% \author{Anonymous Author}

% Shaswata Mitra, Subash Neupane, Trisha Chakraborty, Sudip Mittal, Aritran Piplai, Manas Gaur and Shahram Rahimi

\author{Shaswata Mitra\inst{1}\orcidID{0009-0002-9722-5312}
\and
Subash Neupane\inst{1}\orcidID{0000-0001-9260-3914}
\and
Trisha Chakraborty\inst{1}\orcidID{0009-0002-8531-0667}
\and
Sudip Mittal\inst{1}\orcidID{0000-0001-9151-8347}
\and
Aritran Piplai\inst{2}\orcidID{0000-0002-6437-1324}
\and
Manas Gaur\inst{3}\orcidID{0000-0002-5411-2230}
\and
Shahram Rahimi\inst{1}\orcidID{0000-0003-2779-0076}
}

\authorrunning{S. Mitra et al.}
\institute{Mississippi State University, MS, USA\\ 
\email{\{sm3843, sn922, tc2006\}@msstate.edu,
\{mittal, rahimi\}@cse.msstate.edu}\\
\and
The University of Texas at El Paso, TX, USA\\
\email{apiplai@utep.edu}\\
\and
University of Maryland Baltimore County, MD, USA\\
\email{manas@umbc.edu}
}

\maketitle              % typeset the header of the contribution
\begin{abstract}
    Security Operations Center (SoC) analysts gather threat reports from openly accessible \textit{global threat repositories} and tailor the information to their organization's needs, such as developing threat intelligence and security policies. They also depend on organizational internal repositories, which act as private \textit{local knowledge database}. These local knowledge databases store credible cyber intelligence, critical operational and infrastructure details. SoCs undertake a \textit{manual labor-intensive} task of utilizing these global threat repositories and local knowledge databases to create both organization-specific threat intelligence and mitigation policies. Recently, Large Language Models (LLMs) have shown the capability to process diverse knowledge sources efficiently. We leverage this ability to automate this organization-specific threat intelligence generation. We present \AlgoName, a novel automated threat intelligence contextualization framework that retrieves zero-day vulnerability reports from the global threat repositories and uses its local knowledge database to determine implications and mitigation strategies to alert and assist the SoC analyst. \AlgoName comprises two key phases: \textit{knowledge retrieval} and \textit{contextualization}. Quantitative and qualitative assessment has shown effectiveness in generating up to 93\% accurate organizational threat intelligence with 64\% inter-rater agreement.

\end{abstract}

% ##%%## I miss some notes of related work already in the introduction.
% ##%%## Discussions around additional organizational and analyst workflows that could be implemented given the framework's modular nature.
\section{\SectionIntroduction}

     % There were 2,365 cyberattacks in 2023, marking a 72\% increase in data breaches since 2021 \footnote{https://bit.ly/3zccFKK}. The global cost of cybercrime is estimated to reach \$23.84 trillion by 2027, up from \$8.44 trillion in 2022\footnote{https://bit.ly/476XWNY}. 
     In 2023, there were 2,365 cyberattacks, with 29,065\footnote{URLs: bit.ly/3zccFKK and bit.ly/4g8bdKk} reported Common Vulnerabilities and Exposures (CVE)\footnotemark{}. Cyber analysts in the Security Operations Center (SoC) retrieve malware samples from the internet. These samples are executed in sandboxes for behavior analysis. This analysis leads to developing defensive strategies to detect and prevent cyber-attacks that use such malware. The findings are shared publicly as \textit{generic cyber threat intelligence (CTI)} in \textit{global threat repositories} like CVE, National Vulnerability Database (NVD)\footnotemark[\value{footnote}], Common Weakness Enumeration (CWE)\footnotemark[\value{footnote}] or as third-party threat reports. Security analysts of an organization manually contextualize this generic knowledge to that organization's unique operating conditions by considering factors like network, hardware \& software specifics and business needs to protect from such cyber-attacks. Security measures such as policies and protocols are then deployed depending on this contextualized information to maintain secured operations. Organizations maintain this operating information and contextualized threat intelligence documented in their \textit{local knowledge database}.

    \footnotetext{CVE: cve.mitre.org | CWE: cwe.mitre.org | NVD: nvd.nist.gov}

     % Consider a scenario where a new vulnerability is reported in a global threat repository within a specific process. Parallelly, an organization utilizes the process with operation-specific modifications— documented in its local knowledge database. For example, the modification can be a simple change of IP and port numbers to more complex ones. 
     
     However, expeditiously developing appropriate contextualized reports is a critical challenge before deploying security policies. Manual generation not only consumes high costs but can also be erroneous and require plenty of time due to the volume and criticality of unstructured information. On the other hand, organizations must immediately integrate policies for any novel threat to safeguard operations. Failure of \textit{timely and correct} contextualized CTI generation for policy updation can incur heavy losses. Consider a couple of scenarios where either knowledge's availability is insufficient. \textbf{Scenario 1}: An organization's internal rules detect an unknown process attempting to communicate with an external server. The Endpoint Detection and Response (EDR) team flags and blocks the process. However, without global CTI, they are unaware that this process is part of a larger ransomware campaign. Without this knowledge, the EDR team's response is inadequate, as it fails to recognize additional Indicators of Compromises (IoCs). A secondary payload may go undetected, encrypting the organization's data. \textbf{Scenario 2}: During a routine penetration test, the software and corresponding versions used by the company are identified. Based on CVE/CWE data, the testers flag many software versions due to reported vulnerabilities. However, update log reveals that the flagged software has already been patched, making the alerts unnecessary. Modern IDEs or cybersecurity tools such as Nessus\footnote{Nessus: tenable.com/products/nessus} or Nexpose\footnote{Nexpose: rapid7.com/products/nexpose} can instantly notify the SoC analyst regarding the zero-day vulnerability. However, these solutions cannot suggest accurate counteractions as they cannot assume organizational status since local knowledge resides within the organizational scope. Furthermore, organizations resist granting access to this local knowledge to a third-party vendor. This situation presents a challenge for SoC analysts as they are dealing with two sets of unstructured information. They may require more time to fully understand the context to develop the right policy before the vulnerability gets exploited in an active attack.

     \begin{figure*}
      \includegraphics[width=\textwidth]{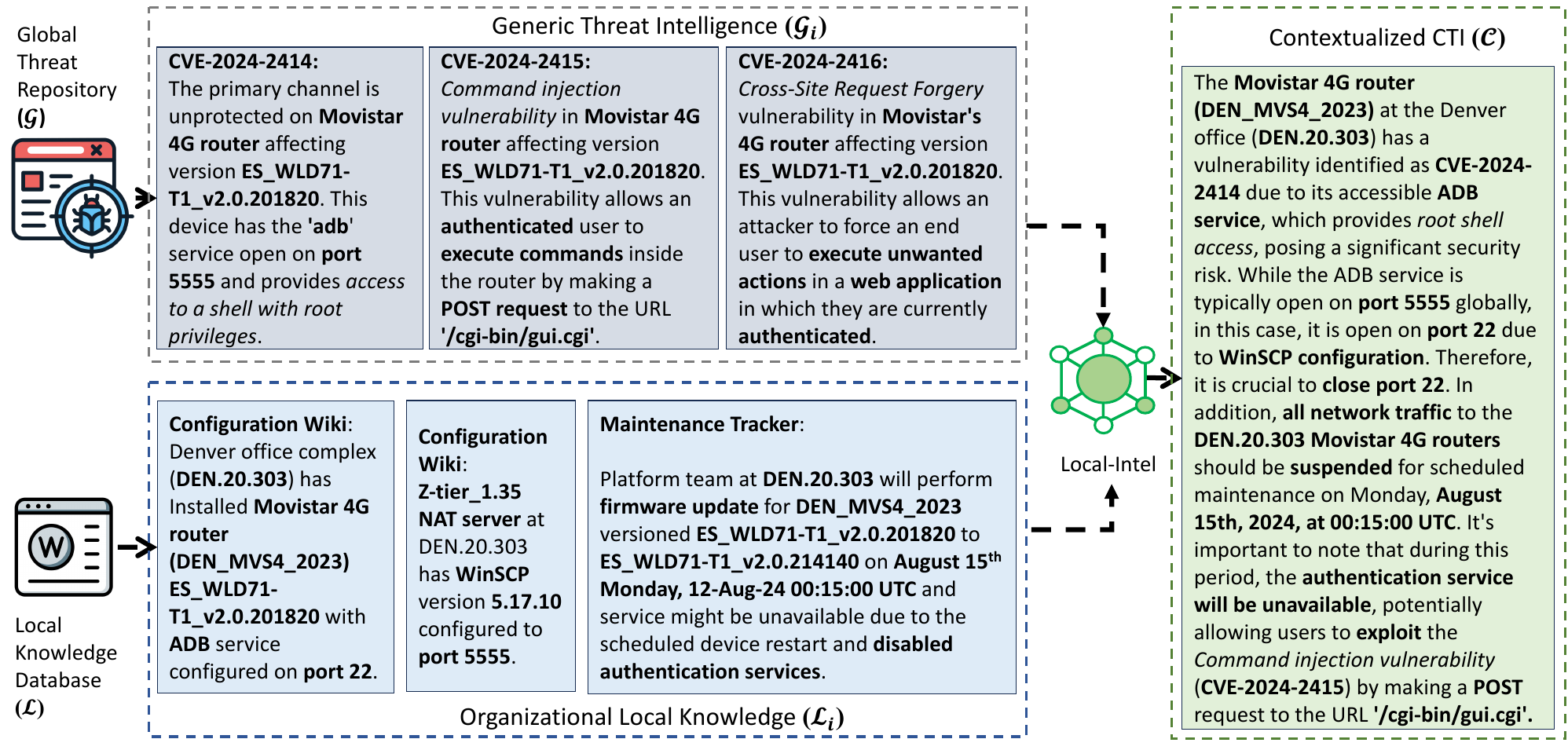}
      \caption{Overview of our \AlgoName framework with an example use case.}
      \label{fig:teaser}
    \end{figure*}
    
     To address this problem, we developed \AlgoName. Our motivation stems from the idea that an \textit{on-premise} system capable of \textit{automatically} generating \textit{relevant} and \textit{accurate} organization-specific threat intelligence, which includes \textit{threat implications} and \textit{counteractions}, by assimilating global and local knowledge, would empower SoC analysts to quickly understand the effects of new cyber threats on their infrastructure, thereby saving valuable time from the manual effort. Hence, SoC analysts can develop, modify, or update their cyber defense strategies in real-time, mitigating the risk of early cyber-attacks. Considering the diverse organizational infrastructure, we have designed our \AlgoName framework to be modular, meaning the framework is customizable based on the use case. To the best of our knowledge, this is the first research that contextualizes global threat intelligence adapted for an organization-specific context. Our work makes the following contributions:

    \begin{itemize}
        \item We demonstrate the feasibility of producing accurate and relevant organization-specific CTI from generic threat intelligence and its operational knowledge. 
        \item We built a knowledge-contextualization framework that generates real-time organizational CTI from publicly available and organizational knowledge.
        \item We construct a prototype repository of local organizational knowledge and an evaluation dataset to assess the generation of contextualized CTI. 
        \item Through our evaluation dataset, we illustrate \AlgoName's ability to generate precise organizational CTI using qualitative and quantitative metrics.
    \end{itemize}

    In Section \ref{problem_and_systemapproach}, we discuss the problem statement and theoretical foundations. Section \ref{system_description} provides a detailed description of our \AlgoName framework. The experiment and evaluation are presented in Section \ref{experiment}. Moving forward, Section \ref{background} explores the related works. Concluding remarks are in Section \ref{conclusion}.

% ############################################################################################################
\section{\SectionObjective}\label{problem_and_systemapproach}% 1 Page

% In this section, we explain the problem definition and theoretical foundations with a series of assumptions associated with the research problem. 

% We commence by defining our research problem. Thereafter, we delve into the proposed solution approach. 
    
    \begin{table}[h]
    \caption{Description of Notation.}
        \begin{center}
        \small
        \begin{tabularx}{0.8\textwidth} { 
          c|>{\raggedright\arraybackslash}X }
         \hline
         \rowcolor{lightgray} 
         \textbf{Notation} & \textbf{Description} \\
        \hline
        % $\prompt$ & Input Prompt \\
        $ \{\glb_i | \glb_i \in \glb\} $ & Global Threat Repository \\
        % $\glb_i : \glb_i \in \glb$ & Retrieved global knowledge from $\glb$ \\
        $ \{\loc_i | \loc_i \in \loc\} $ & Local Organizational Database \\
        % $l : l \in L$ & Retrieved local knowledge from $L$ \\
        $\query$ & Query to fetch global ($\glb_i$) \& local ($\glb$) knowledge  \\
        % $\query_\loc$ & Query to fetch local knowledge $\loc_i$ from $\loc$ \\
        $\completion$ & Contextualized Completion \\
        \hline
        \end{tabularx}
        \end{center}
    \end{table}
    
    % ##%%## Now we have the reverse problem since this section seems a bit cut off from the rest of the paper (esp. 2.1).
    % \subsection{Problem Definition} 
    Global threat repository ($\glb = \{ \glb_1,\glb_2, ..., \glb_n \} $) is a publicly available set of online CTI reports ($\glb_i$). Local knowledge database ($\loc = \{ \loc_1, \loc_2, ..., \loc_n \}$) consists of policies and procedures of an organization's operating environments ($\loc_i$), such as business requirements, trusted cyber intelligence, allowed system software list and version details, cyber knowledge about the organization, asset location and configurations, DMZ configurations, and maintenance reports.  
    
    \begin{boxEnv}
        For a given set of vulnerability $\glb_i$ in $\glb$ and corresponding relevant organizational knowledge $\loc_i$ in $\loc$. The task is to generate Completion ($\completion$), which is the contextualized knowledge of $\glb_i$ w.r.t. $\loc_i$. $\completion$ can be considered as contextualization function $f(\cdot)$, that translates $\glb_i$ to organizational context using $\loc_i$.
        \begin{equation}
            f(\glb_i, \loc_i) = \completion_i  \forall  \glb_i \cap \loc_i \neq \phi  
        \end{equation}
    \end{boxEnv}
    
    For instance, in Figure~\ref{fig:teaser}, where $\glb_i$ is a set containing information stating vulnerability ($v$) through process ($p$). Alternatively, $\loc_i$ contains information regarding the organization using process ($p$) for its operations and other relevant information. Hence, in the process of generating contextualized threat intelligence $\completion$ considering $v$ and $p$, $\glb_i$ is being translated through $\loc_i$, when $\glb_i \cap \loc_i \neq \phi$. 
    
    % For instance, if there is a new malware ($m$) that exploits a vulnerability ($v$) through a process ($p$) that is reported in $\glb$ and an organization uses software ($s$) for its operations, that uses the same process ($p$), which is documented in $\loc$. Hence, to generate a completion $\completion$ for prompt $\prompt$ regarding $m$, relevant knowledge in the global CTI repository $\glb_i$ and local knowledge database $\loc_i$ is required. Therefore, through Completion $\completion$, the knowledge in $\glb_i$ is being contextualized depending upon $\prompt$ and $\loc_i$.
    
    % \noindent The following section will discuss our research background with a literature review.

% ############################################################################################################

% ##%%## The “tool” term is a bit misleading. Fig. 3 is only briefly referred to but never explained in the main text.
\section{\AlgoName ~\SectionSystemDesc}
    \label{system_description} % 2.5 Page   

    In this section, we explain our \AlgoName framework. We first explain our solution and each module with its functionality in detail (refer to Figure~\ref{Fig:research_diag}). Finally, we discuss the system implementation and module interactions to generate the final contextualized threat intelligence $\completion$.
    \vspace{-3mm}

    \subsection{Solution Approach}
    
\begin{figure*}[ht]
          \includegraphics[width=1\textwidth]{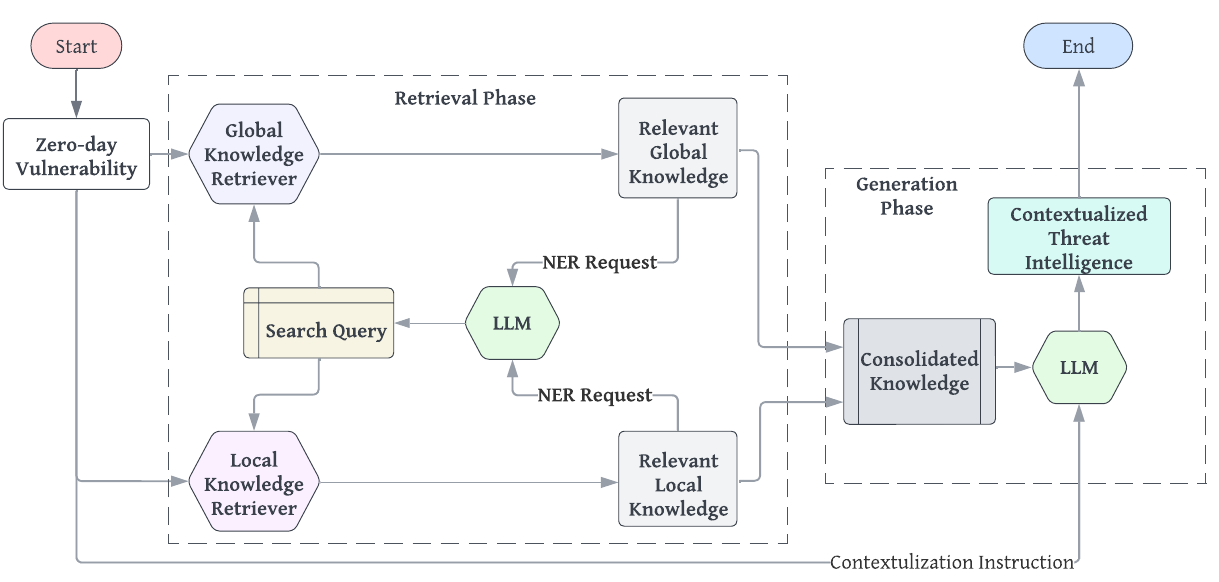}
          \caption{\AlgoName data-flow diagram. The zero-day vulnerability triggers the system to retrieve information from global $\glb_i$ and local $\loc_i$ knowledge sources and then contextualizes the results, producing a final output Completion $\completion$.} 
          \label{Fig:process_flow}
        \end{figure*}

        \AlgoName consists of two core phases: knowledge retrieval \textit{(Retrieval Phase)} and generation \textit{(Generation Phase)}. In the \textit{retrieval phase}, knowledge from \textit{global ($\glb$)} and \textit{local ($\loc$)} sources are retrieved, and in the \textit{generation phase}, a final contextualized threat intelligence $\completion$ based on the retrieved knowledge $\glb_i \cup \loc_i$ is generated. Refer to Figure \ref{Fig:process_flow} and Algorithm~\ref{alg:solution_approach} for framework overview.
    
        \vspace{-5mm}
        
        \begin{algorithm}[!h]
            % \small
            \caption{\AlgoName Pseudo-code} \label{alg:solution_approach}
            \KwInput{Generic Threat Intelligence ($\glb_i$)} 
            \KwOutputput{Contextualized Threat Intelligence ($\completion$)}
            \textsc{\textbf{Retrieval Phase:}} \\
            % $\query \leftarrow get\_search\_query(\glb_i)$ \\
            $\loc_i \leftarrow execute\_local\_search (\glb_i, \loc)$ \\
            \While{ $\loc_i \cap \overline{\glb_i}  \neq \phi$ }{
                $\query \leftarrow get\_search\_query(\glb_i \cup \loc_i)$ \\
                \ForAll{$\alpha \in \query$}{
                    $\glb_i \leftarrow execute\_global\_search (\alpha, \glb)$ \\
                }
                $\loc_i \leftarrow execute\_local\_search (\glb_i, \loc)$ \\
                % \ForAll{$\beta \in \query$}{
                %     $\loc_i \leftarrow execute\_local\_search (\glb_i, \loc)$ \\
                % }
            }
            
            \textsc{\textbf{Generation Phase:}}\\
            $\completion \leftarrow generate\_completion(\glb_i \cup \loc_i) $ \\
            return $\completion$
    
            % $\query_\loc \leftarrow get\_local\_search\_keywords(\prompt \cup \glb_i)$ \\
        \end{algorithm}
        \vspace{-5mm}
        
        \begin{itemize} % declare phase consistency
            \item In the Retrieval Phase, the system retrieves generic CTI $\glb_i$ from $\glb$ and relevant local knowledge $\loc_i$ from $\loc$ based on the relevancy. The system performs Named Entity Recognition (NER) to identify search keywords/queries $\query$ over on the acquired knowledge ($\glb_i \cup \loc_i$). Then, it executes the search for all search queries in $\query$ in the global knowledge repository $\glb$ and local knowledge database $\loc$ to fetch relevant threat reports and associated details. This phase continues until no additional knowledge is required ($ \loc \cap \overline{\glb_i} \neq \phi $) to generate final contextualized threat intelligence.
            
            % \item In Phase 2, the system retrieves relevant local knowledge $\loc_i$ from $\loc$ based on the relevancy of $\glb_i$ and $\prompt$. The system generates search query $\query_\loc$ using the input Prompt combined with retrieved global knowledge ($\prompt \cup \glb_i$) to achieve this. $\query_\loc$ is then executed in $\loc$ and $\loc_i$ is retrieved.
            
            \item Finally, in the Generation Phase, the system generates contextualized threat intelligence $\completion$ for the zero-day vulnerability generic threat intelligence based on the retrieved global knowledge and local knowledge ($\glb_i \cup \loc_i$).
        \end{itemize}

    \subsection{\AlgoName System Modules}
        To implement \AlgoName, we first discuss system modules, which are \textit{Global Threat Repository ($\glb$), Local Knowledge Database ($\loc$), Agent, Tools}, \textit{LLM}, with zero-day threat report input $\glb_i$ and contextualized completion $\completion$) output.

        \subsubsection{Global Threat Repository ($\glb$)} refers to publicly available cybersecurity threat intelligence (CTI), such as threat reports from CVE, NVD, CWE, security blogs and bulletins, social media updates, and third-party reports. These repositories contain well-documented reports on cybersecurity threats, such as malware, vulnerability, cyber attacks, and many more. The primary purpose of these repositories is to facilitate information sharing among cybersecurity professionals regarding the latest developments. However, the global knowledge $\glb_i$ is generic and may not directly apply to an organization's needs as organizations tend to customize their infrastructure depending upon the business. Moreover, the knowledge obtained from unverifiable sources is only directly usable by an organization after thorough analysis. \AlgoName is expected to be connected to these threat repositories for automated zero-day vulnerability report retrieval.

        % \aritran{We need a sentence here saying that global knowledge is too generic, and may not be directly applicable to a local organization's needs. Moreover, the knowledge derived from unverifiable sources such as Wikipedia is not usable by a local organization. }

        % \manas{The reference to Confluence and Notion is very generic. Can we provide an appropriate reference, such as a wiki made using notion by any organization? Also, should we write wikis or wikis?}

        % \aritran {Add a sentence here: global knowledge contains a wide-range of CTI, but they need to be supplanted with either local organization-specific information, or trusted knowledge preserved by the same organization. Also, can we change the term to Local Database or Local Trusted Database?} 

        \subsubsection{Local Knowledge Database ($\loc$)} refers to an organizations' operational information repository. Due to the generic characteristics, $\glb$ contains a wide range of CTI, but they must be supplanted with organization-specific information to be useable. Hence, local knowledge databases or wikis are private knowledge repositories containing critical information related to organizational operations and trusted threat intelligence, such as specifics regarding the environment, operating systems, infrastructure, software, third-party systems, and processes. Confluence\footnote{Confluence: atlassian.com/software}, Notion\footnote{Notion: notion.so/product/wikis}, are a few instances of such wiki platforms. The primary goal of these wikis is to facilitate structured development and knowledge sharing among the working professionals in an organization. Due to the unstructured nature of this information, we assume wiki platforms to be our local knowledge database. However, more structured sources like knowledge graphs can replace them with similar searching functionalities.

        %\footnote{VD, Website:learn.microsoft.com/en-us/} %\sudip{why are we crediting microsoft for VD? these have been around for decades, you can remove this footnote} 
        % \subsubsection{Vector Database} A Vector Database (VD) is a specific type of database that stores data in the form of high-dimensional vectors. These vectors are mathematical representations derived from raw data, such as unstructured texts, while representing word-level semantic meaning. The advantage of a vector database is that it facilitates rapid similarity search and retrieval of data based on their vector distances, contrary to traditional databases that search based on exact matches or predefined criteria. In our \AlgoName framework, Chroma DB\footnote{Chroma: trychroma.com}, an in-memory VD, allows storing and retrieving the most relevant pre-processed local organizational knowledge based on the semantic meaning of the prompt.

        % \subsubsection{Embedding Model} Embedding refers to the technique used for representing natural language words and documents in a way that captures their meanings. This representation is typically a real-valued vector in low-dimensional space. In our \AlgoName framework, the embedding model is used to vectorize and pre-process the local knowledge databases $\loc$ to store them in a vector database. This allows performing a \textit{semantic search} for relevant knowledge retrieval.        

        \subsubsection{Agent} is the main controller in our \AlgoName framework. It controls the overall flow, from receiving the input vulnerability report trigger to returning the final contextualized completion $\completion$. Specifically, the Agent's function is to determine and regulate the sequence of actions among two phases for generating the output. The Agent actions are primarily of three types: \textit{Query generation}, \textit{Query execution}, and \textit{Completion generation}. To achieve this, the Agent interacts with the other two modules: \textit{Tool} and \textit{LLM}, detailed following. 

        \begin{itemize}
            \item \textit{Query generation} refers to generating search queries for information retrieval from either $\glb$ or $\loc$. The Agent generates a search query to retrieve all the relevant information from pre-acquired knowledge ($\glb_i \cup \loc_i$). The Agent performs contextual embedding and keyword identification through named entity recognition (NER) to generate search queries $\query$.
            
            \item \textit{Query execution} refers to executing search query $\query$ in global threat repository $\glb$ and local knowledge database $\loc$ to retrieve relevant knowledge. Due to the different characteristics of $\glb$ and $\loc$, the retrieval process can either be a keyword search through online API calls or a semantic similarity search.
            
            \item \textit{Text generation} can be considered an LLM inference scenario, where the Agent passes an input text to generate desired output text using LLM. Task-specific input prompts are pre-designed in the Agent. 
        \end{itemize}

        \subsubsection{Tool} are functions that help the Agent execute some third-party actions. The actions can be diverse in type, for instance, making an online API call, performing a database search, executing custom scripts, invoking other software, and many more. However, for the scope of our research, tool functionality is limited to query generation using LLM, query execution through API calls and vector database search, and contextualized generation using LLM functionalities only. Therefore, in our \AlgoName framework, tools are responsible for executing online searches through API calls or vector database searches and parsing the results while bridging the Agent's access to different framework modules.

        \subsubsection{Large Language Model (LLM)} acts as the brain of our \AlgoName framework to process diverse information and generate contextualized CTI. Besides contextualized threat intelligence generation, it acts as the parser that processes retrieved knowledge to generate queries for structured information retrieval. Depending on the task, the Agent invokes LLM with instructions and information.

        \subsubsection{Input: Zero-day threat report ($\glb_i$)} refers to publicly available CTI reports regarding any discovered vulnerability or malware. We assume that \AlgoName is connected with the global CTI repositories ($\glb$) with active triggers to receive any newly disclosed threat reports for instant processing.

        \subsubsection{Output: Contextualized completion ($\completion$)} is the real-time generated threat intelligence specifically tailored for an organization depending on its unique operating condition. The objective of $\completion$ is to assist SoC by providing mitigating strategies or relevant information on the specific zero-day threat ($\glb_i$). We assume the local knowledge database ($\loc$) contains all required organizational knowledge.

    % ##%%## none of these prompts that are engineered are ever presented to the readers and the complexity involved in presenting it is never shared with the readers
    \subsection{\AlgoName Implementation \& Module Interactions}

        \begin{figure*}[!ht]
          \centering
          \includegraphics[width=1\textwidth]{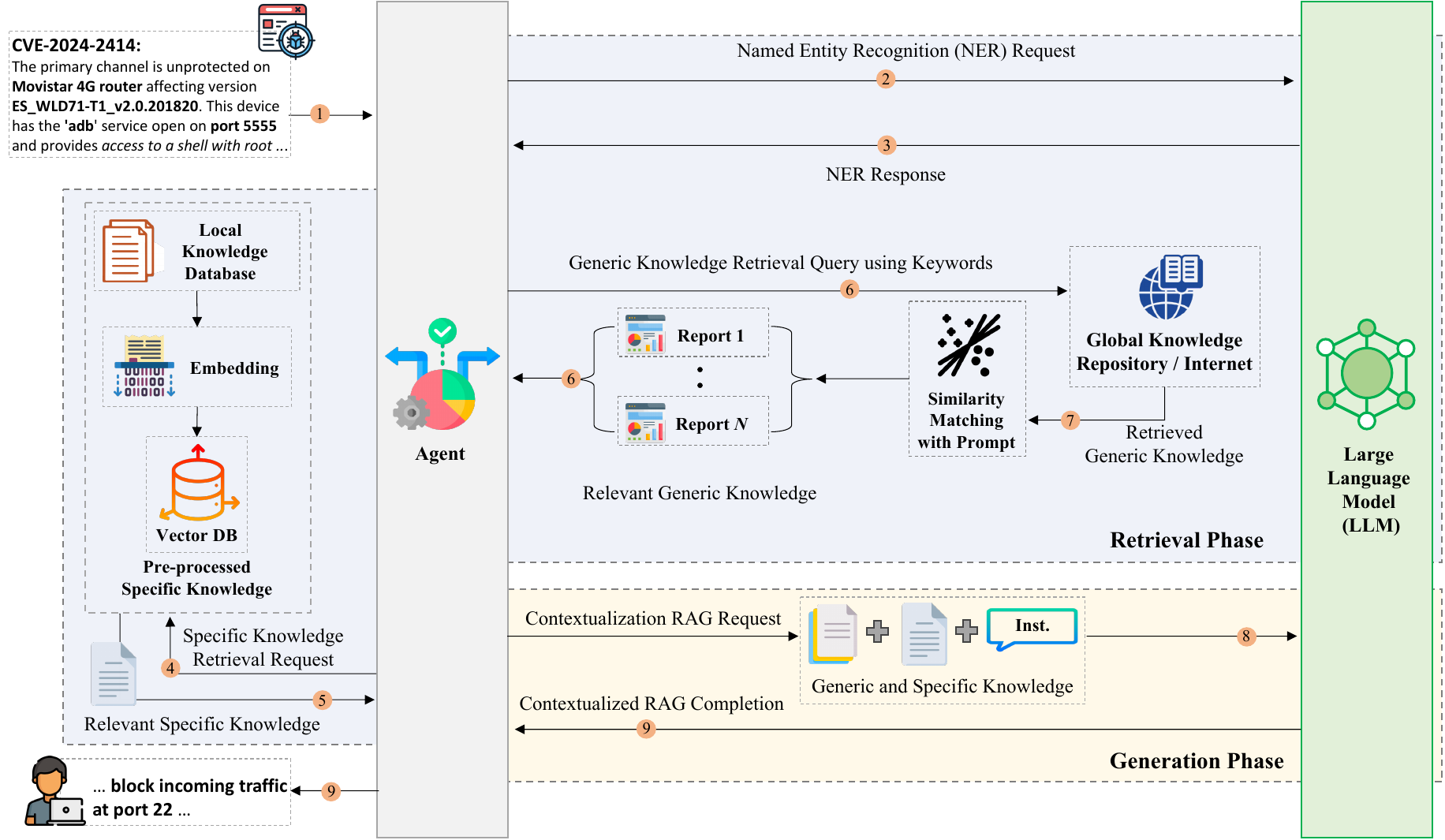}
          \caption{\AlgoName framework module interaction in the two phases: knowledge retrieval, and contextualized threat intelligence generation. Processes are numbered in ascending execution order following the data-flow diagram (Fig. \ref{Fig:process_flow}). }
          \label{Fig:research_diag}
        \end{figure*}

        Previously, we have described each module in the architecture. Here, we explain the implementation phases with intermediate module interactions (refer to Figure~\ref{Fig:research_diag}). \AlgoName initiates when a vulnerability report is received. The report can be pushed manually or via automated zero-day triggers.

        \subsubsection{Knowledge Retrieval (Phase 1):}
        This is the first phase of our framework where the \textit{Agent} retrieves generic threat intelligence ($\glb_i$). It generates search queries ($\query$) for relevant knowledge retrieval. The initial local knowledge search (refer to Algorithm \ref{alg:solution_approach}) plays a crucial role in identifying whether the threat intelligence is relevant to the organization. If there is no overlap ($\glb_i \cap \loc = \phi$), then the Agent discards input $\glb_i$, as there are no connections; hence, it cannot be contextualized. Upon overlaps discovered, it iteratively generates $\query$ and executes knowledge retrieval from both global ($\glb$) and local ($\loc$) sources until all required knowledge needed to be considered for contextualization is retrieved. For the scope of our experiment, we implemented global knowledge retrieval from the Internet through keyword search via API endpoints of global threat repositories such as NIST, CVE, and ensemble \cite{arabzadeh2021predicting} vector similarity search for local knowledge retrieval from the organizational wikis. This simplified approach efficiently fetches corresponding relevant knowledge from both sources. For instance, for the following threat intelligence, the execution is as follows:

        \begin{tcolorbox}[enhanced,attach boxed title to top center={yshift=-3mm,yshifttext=-1mm},
            colback=white,colframe=gray!75!black,colbacktitle=gray!80!black, title=Invoked Generic Threat Intelligence ($\glb_i$),
              boxed title style={size=small,colframe=gray!90!black}, left=0.5mm, right=0.5mm, boxrule=0.75pt ]
              %boxrule=0.0pt,
              %boxsep=2pt 
              \small
              \textbf{CVE-2024-2414}: The primary channel is unprotected on Movistar 4G router affecting E version S\_WLD71-T1\_v2.0.201820. This device has the `adb' service open on port 5555 and provides access to a shell with root privileges.
        \end{tcolorbox}

        Upon receiving $\glb_i$ above, the Agent generates query embedding to perform ensemble retrieval in $\loc$. After executing $\query$ in the vector-indexed $\loc$, the Agent identifies ``Movistar 4G" to be the affecting device with following knowledge:

        \begin{tcolorbox}[enhanced,attach boxed title to top center={yshift=-3mm,yshifttext=-1mm},
            colback=white,colframe=gray!75!black,colbacktitle=gray!80!black, title=Phase 1: NER Query Generation ($\query$),
              boxed title style={size=small,colframe=gray!90!black}, left=0.5mm, right=0.5mm, boxrule=0.75pt ]
              %boxrule=0.0pt,
              %boxsep=2pt 
              \small
              \textbf{Agent Instruction}:\\
              \textcolor{teal}{
                You are a named entity recognition (NER) tool. Given the following classes, perform NER for the provided Input text.\\
                Classes: software, device, library, functionality, attack\_vector, vulnerability ...}\\
              \textbf{Input Threat Intelligence}:\\
              \textcolor{green!45!black}{
                  CVE-2024-2414: The primary channel is unprotected on Movistar 4G router affecting E version S\_WLD71-T1\_v2.0.201820. This device has the `adb' service open on port 5555 and provides access to a shell with root privileges.
                }\\
              \textbf{Output Keywords}:\\
              \textcolor{black}{
              \{"device": ``Movistar 4G'', "attack\_vector": ``port 5555'', "functionality": ``adb'' \}
              }
        \end{tcolorbox}

        The semantic search is performed through vector embedding generation and execution of similarity matching algorithms (cosine, euclidean, dot-product).

        \begin{tcolorbox}[enhanced,attach boxed title to top center={yshift=-3mm,yshifttext=-1mm},
            colback=white,colframe=gray!75!black,colbacktitle=gray!80!black, title=Phase 1: Retrieved Local Knowledge ($\loc_i$) using $\query_\loc$,
              boxed title style={size=small,colframe=gray!90!black}, left=0.5mm, right=0.5mm, boxrule=0.75pt ]
              %boxrule=0.0pt,
              %boxsep=2pt 
              \small
              \textbf{Configuration Wiki:}:\\
              \textcolor{teal}{
                -- Denver office complex (DEN.20.303) has Installed \textbf{Movistar 4G router} (DEN\_MVS4\_2023) \textbf{ES\_WLD71-T1\_v2.0.201820} with ADB service configured on port 22.\\
                -- Z-tier\_1.35 NAT server at DEN.20.303 has WinSCP version 5.17.10 configured to \textbf{port 5555}.
            }
        \end{tcolorbox}

        After searching $\loc$, the Agent performs similar query generation $\query$ and execution iteratively in $\glb$ and $\loc$ for additional context retrieval. For this example, the additional retrieved knowledge $\glb_i$ and $\loc_i$ from Phase 1 is below:

        \begin{tcolorbox}[enhanced,attach boxed title to top center={yshift=-3mm,yshifttext=-1mm},
            colback=white!10!white,colframe=gray!75!black,colbacktitle=gray!80!black,
            title=Phase 1: Additional Global and Local Knowledge ($\glb_i \cup \loc_i$),
            boxed title style={size=small,colframe=gray!50!black}, left=0.5mm, right=0.5mm, boxrule=0.75pt]
            \small
           
            \textcolor{black}{\textbf{Global Knowledge}}:\\
            \textcolor{teal}{
            -- \textbf{CVE-2024-2415}: Command injection vulnerability in \textbf{Movistar 4G router} affecting version \textbf{ES\_WLD71-T1\_v2.0.201820}. This vulnerability allows an authenticated user to execute commands inside the router by making a POST request to the URL \textbf{'/cgi-bin/gui.cgi'}.\\
            -- \textbf{CVE-2024-2416}: Cross-Site Request Forgery vulnerability in Movistar's 4G router affecting version \textbf{ES\_WLD71-T1\_v2.0.201820}. This vulnerability allows an attacker to force an end user to execute unwanted actions in a web application in which they are currently authenticated.
            }\\
            \textcolor{black}{\textbf{Local Knowledge}: \\ }\textcolor{green!45!black}{
            -- \textbf{Maintenance Tracker}: Platform team at \textbf{DEN.20.303} will perform firmware update for \textbf{DEN\_MVS4\_2023} versioned \textbf{ES\_WLD71-T1\_v2.0.201820} to ES\_WLD71-T1\_v2.0.214140 on August 15th Monday, 12-Aug-24 00:15:00 UTC and service might be unavailable due to the scheduled device restart and disabled authentication services.            } 
        
            % $>$ \textbf{\textcolor{black}{Finished}}
        \end{tcolorbox}

        Consolidated knowledge ($\glb_i \cup \loc_i$) is passed for query generation. For this case, \textit{DEN.20.303} and \textit{DEN\_MVS4\_2023} used to identify related information.
        
        \subsubsection{Contextualized Generation (Phase 2):}
        In this phase, upon complete retrieval of $\glb_i \cup \loc_i$, the Agent invokes LLM for final contextualization. 

        % The pre-processed embedded local knowledge database $\loc$ is stored in the Chroma DB to accomplish this. Therefore, to retrieve relevant local knowledge $\loc_i$, the Agent generates queries $\query_\loc$ through LLM based on the Prompt and retrieved global knowledge $\prompt \cup \glb_i$. 

        \begin{tcolorbox}[enhanced,attach boxed title to top center={yshift=-3mm,yshifttext=-1mm},
            colback=white,colframe=gray!75!black,colbacktitle=gray!80!black, title=Phase 2: Contextualized Organizational Threat Intelligence ($\completion$),
              boxed title style={size=small,colframe=gray!90!black}, left=0.5mm, right=0.5mm, boxrule=0.75pt ]
              %boxrule=0.0pt,
              %boxsep=2pt 
              \small
              \textcolor{black}{\textbf{Agent Instruction}}:\\ 
              \textcolor{teal}{You are an honest network security analyst. Given public threat intelligence reports fetched from trusted cybersecurity sources and organizational infrastructure and operations details. Generate a cyber threat intelligence report with all details, including the impact and mitigation strategies. Do not include any information that is not provided as additional knowledge.}\\
              \textcolor{black}{\textbf{Retrieved Global Knowledge ($\glb_i$)}}:\\ 
              \textcolor{green!45!black}{
               -- \textbf{CVE-2024-2414}: The primary channel is unprotected on Movistar 4G router affecting E version S\_WLD71-T1\_v2.0.201820. This device has the `adb' service open on port 5555 and provides access to a shell with root privileges.\\
              -- \textbf{CVE-2024-2415}: Command injection vulnerability in Movistar 4G router affecting version ES\_WLD71-T1\_v2.0.201820. This vulnerability allows an authenticated user to execute commands inside the router by making a POST request to the URL '/cgi-bin/gui.cgi'.\\
              -- \textbf{CVE-2024-2416}: Cross-Site Request Forgery vulnerability in Movistar's 4G router affecting version ES\_WLD71-T1\_v2.0.201820. This vulnerability allows an attacker to force an end user to execute unwanted actions in a web application in which they are currently authenticated.
              }\\
              \textcolor{black}{\textbf{Retrieved Local Knowledge ($\loc_i$)}}:\\ 
              \textcolor{green!45!black}{
              -- \textbf{Configuration Wiki}: Denver office complex (DEN.20.303) has Installed Movistar 4G router (DEN\_MVS4\_2023) ES\_WLD71-T1\_v2.0.201820 with ADB service configured on port 22\\
              -- \textbf{Configuration Wiki}: Z-tier\_1.35 NAT server at DEN.20.303 has WinSCP version 5.17.10 configured to port 5555.\\
              -- \textbf{Maintenance Tracker}: Platform team at DEN.20.303 will perform firmware update for DEN\_MVS4\_2023 versioned ES\_WLD71-T1\_v2.0.201820 to ES\_WLD71-T1\_v2.0.214140 on August 15th Monday, 12-Aug-24 00:15:00 UTC and service might be unavailable due to the scheduled device restart and disabled authentication services.}\\
              \textcolor{black}{\textbf{Contextualized Completion}}:\\
              \textcolor{black}{The \textbf{Movistar 4G router (DEN\_MVS4\_2023)} at the \textbf{Denver office (DEN.20.303)} has a vulnerability identified as \textbf{CVE-2024-2414} due to its accessible ADB service, which provides root shell access, posing a significant security risk. \textbf{While the ADB service is typically open on port 5555 globally, in this case, it is open on port 22 due to WinSCP configuration}. Therefore, it is crucial to \textbf{close port 22}. In addition, all \textbf{network traffic to the DEN.20.303 Movistar 4G routers should be suspended} for scheduled maintenance on \textbf{Monday, August 15th, 2024, at 00:15:00 UTC}. It's important to note that during this period, the \textbf{authentication service will be unavailable, potentially allowing users to exploit the Command injection vulnerability (CVE-2024-2415)} by making a POST request to the URL '/cgi-bin/gui.cgi'.}
        \end{tcolorbox}

        After completing both phases, we can observe that through an initial local search, the agent identified the ``Movistar 4G router'' as the device of interest ($\glb_i \cap \loc$) with additional relevant knowledge. Then, it iteratively retrieved additional threat intelligence (CVE-2024-2415 and CVE-2024-2416) for the device from $\glb$ (we considered NVD as $\glb$ for the experiment) and $\loc$ (considered an organizational wiki) to obtain additional local context. The concatenation of knowledge prior to retrieval allowed the discovery of indirect relevant documents such as maintenance schedules. Without this knowledge, the mitigation strategy might become ineffective. Finally, by providing all relevant information ($\glb_i \cup \loc_i$) and task instruction, the Agent invokes LLM for real-time organization-specific threat intelligence generation. This relevant real-time update then equips SoC analysts with all relevant information without investing any time in manual investigation. The SoC analyst can then utilize this knowledge to take the necessary actions to safeguard the organization against imminent cyber threats.

\section{\SectionExperiment}\label{experiment}% 1.5 Page

In this section, we discuss our experiments and the achieved evaluation results. For our evaluations, we performed experiments considering 58 publicly available threat intelligence scenarios to demonstrate the feasibility of the \AlgoName framework and assess contextualization relevancy. For the global threat repository ($\glb$), we considered NVD-CVE data, and for the local knowledge database ($\loc$), a curated organizational wiki (PII anonymized for confidentiality). However, as described in Section \ref{system_description}, \AlgoName \footnote{LocalIntel Repository: github.com/shaswata09/LocalIntel} is modular, allowing flexibility to modify the modules depending on requirements and organization-specifics. For example, other generic threat intelligence sources can be integrated with $\glb$, different local knowledge sources such as knowledge graphs can be incorporated, and other generative language models can be adopted for a more controlled generation. Following, we will delve into the evaluation dataset and experiment setup. Finally, we will describe our evaluation measures and findings with justifications.

    \subsection{Data Description and Experiment Setup}
    
    Our dataset includes (1) \textbf{58} trigger/zero-day generic threat intelligence reports ($\glb_i$), (2) \textbf{5 organizational wikis} resembling an organizational local knowledge database source ($\loc$), and (3) \textbf{58} subject matter expert (SME) generated (manually unbiased) ground truth ($\overline{\completion}$). A trigger ($\glb_i$) can be a report of any malware, vulnerability, attack vector, or security updates. For further automated relevant global knowledge retrieval, \AlgoName is connected with CVE API endpoints. The 58 trigger reports contain both positive and negative test cases. We gathered \textbf{5 organizational wikis} corresponding to 5 real-time applications and curated them (PII removed) suitable to the research. For each positive test scenario, we ensured the corresponding knowledge was present in the local knowledge database. In addition to the organizational wiki, we also collected \textbf{326} confidential organizational trusted CTI reports to allow \AlgoName to retrieve more infrastructural context and threat implications. These reports offer detailed analyses and insights from security analysts studying various global cyber attacks within the organization. For negative test scenarios, there was no intersecting knowledge present in $\loc$ i.e. $\glb_i \cap \loc = \phi$. 
    In conducting our experiments, we tested with proprietary \textit{GPT-3.5-turbo, and GPT-4o\footnote{GPT Models: platform.openai.com/docs}, and open-source meta-llama/Llama-2-7b-chat-hf, meta-llama/Meta-Llama-3.1-8B-Instruct, mistralai/Mistral-7B-Instruct-v0.2, nvidia/Mistral-NeMo-Minitron-8B-Base, Qwen/Qwen1.5-7B-Chat, AiMavenAi/AiMaven-Prometheus, senseable /WestLake-7B-v2, PetroGPT/WestSeverus-7B-DPO-v2} downloaded from \textit{huggingface.co} as the LLM models. All models' temperatures were deliberately kept default, and instructions prompts were set the same for neutral comparison. The global knowledge was retrieved from NVD-CVE sources through search API. For local knowledge retrieval, we store the 5 organizational wikis and 326 threat reports in a vector database (Chroma \footnote{Chroma: trychroma.com}). We segmented and organized the data into smaller chunks to enhance processing efficiency. In our experimental setup, we opted for a \textit{chunk size} of 1500 with a \textit{chunk overlap} of 150. We used the \textit{text-embedding-ada-002}\footnote{OpenAI Embedding: platform.openai.com/docs} as our base model for embedding each chunk of data in Chroma DB and used Maximal Marginal Relevance (MMR) sorting for dense retrieval of relevant chunks. The experiment was performed over Intel i9-12900 with 24 GB GeForce RTX™ 3090Ti GPU and 128 GB of RAM.

    \begin{table}[]
    \centering
    \footnotesize
    \caption{Evaluation results of our \AlgoName framework over following LLMs.\\}\label{Table:results}
    \renewcommand{\arraystretch}{1.3}
    \begin{tabular}{p{4cm}| >{\raggedleft\arraybackslash}p{2.2cm}|>{\raggedleft\arraybackslash} p{2.2cm}|>{\raggedleft\arraybackslash}p{2.2cm}}
    \hline
        \rowcolor{lightgray} 
        \textbf{Model} & \textbf{Ragas (Sim.)} & \textbf{GEval (Cor.)} & \textbf{BertSc-F1}  \\ \hline
        
        gpt-3.5-turbo & 0.92 & 0.75 & 0.68 \\ \cline{2-4}
        gpt-4o & 0.91 & 0.75 & 0.66 \\ \cline{2-4}
        % palm2 & -- & -- \\ \cline{2-3}
        qwen1.5-7b-chat & 0.92 & 0.78 & 0.66 \\ \cline{2-4}
        llama-3.1-8b-Instruct & 0.85 & 0.46 & 0.53 \\ \cline{2-4}
        westlake-7b-v2 & 0.92 & 0.69 & 0.65 \\ \cline{2-4}
        llama2-7b-chat & 0.91 & 0.69 & 0.65  \\ \cline{2-4}
        mistral-7b-instruct-v2 & 0.90 & 0.67 & 0.63  \\ \cline{2-4}
        prometheus-7b & 0.93 & 0.71 & 0.66  \\ \cline{2-4}
        westseverus-7b-dpo-v2 & 0.90 & 0.60 & 0.60  \\ \cline{2-4}
        mistral-nemo-minitron-8b & 0.84 & 0.56 & 0.55 \\ \hline
        \end{tabular}
    \end{table}

    \begin{figure}[]
        \centering
         \includegraphics[width=0.8\linewidth]{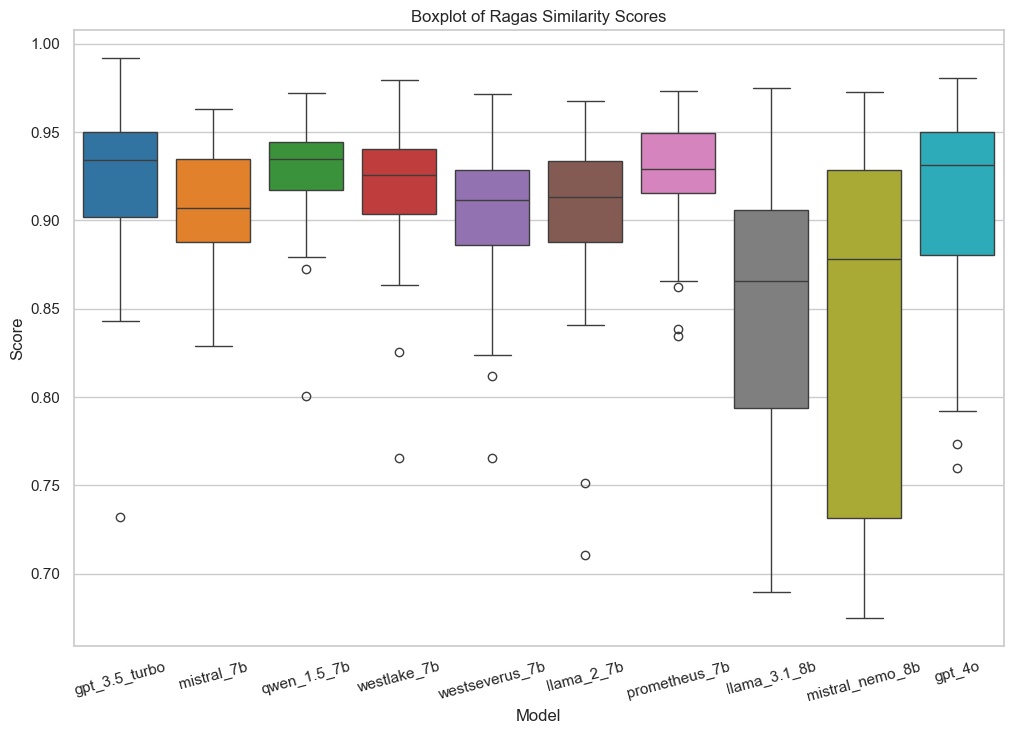}
         \includegraphics[width=0.8\linewidth]{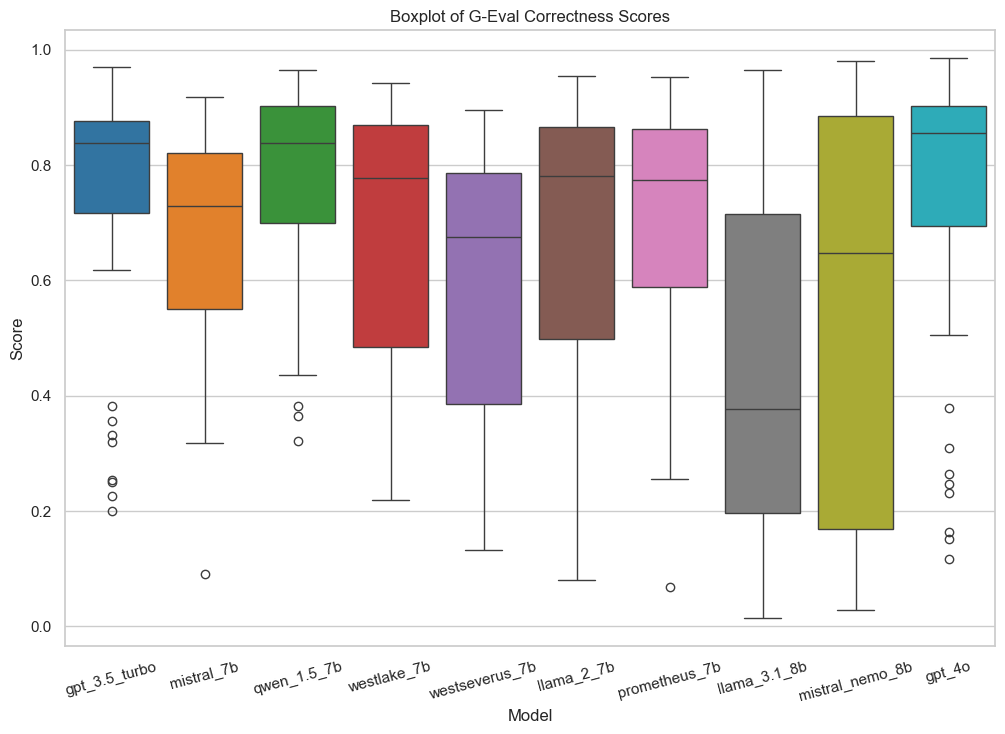}
         \includegraphics[width=0.8\linewidth]{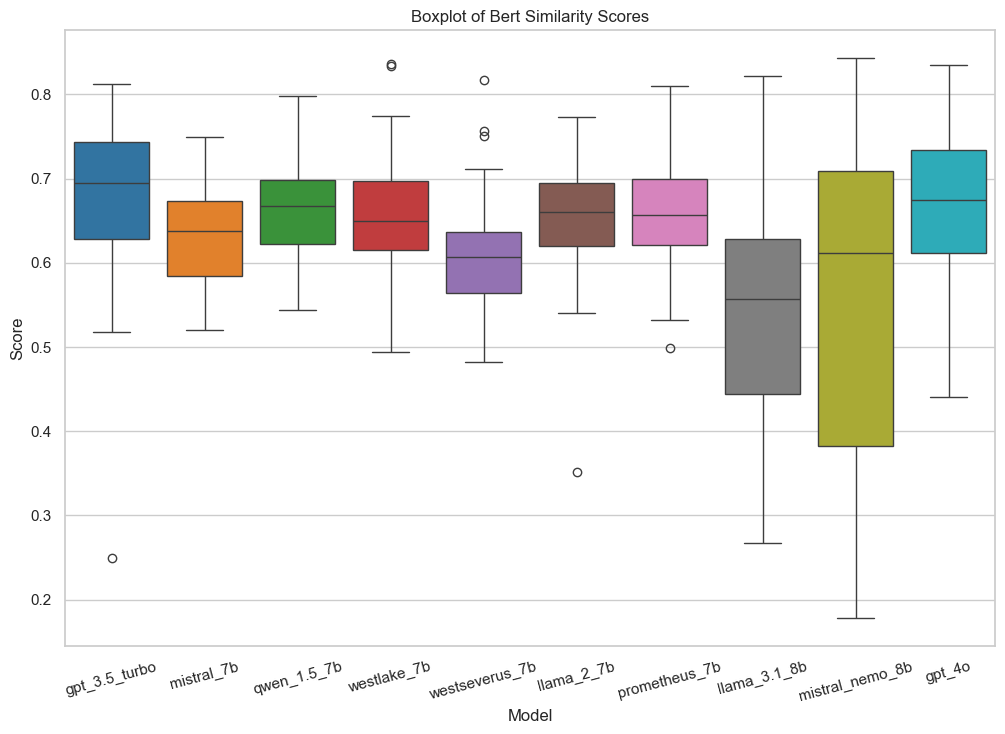}
        \caption[]{Evaluation box-plot for Completion $\completion$ with respect to ground truth.}
        \label{fig:evaluation_plot}
    \end{figure}
    
    \subsection{Quantitative Evaluation}\label{subsec:quantitative_eval} 
        For the quantitative assessment of \AlgoName's performance in generating contextually relevant organizational threat intelligence $\completion$, we utilize three frameworks: Retrieval Augmented Generation Assessment (RAGAs)~\cite{es2023ragas}, G-EVAL~\cite{liu2023g}, and BertScore~\cite{zhang2019bertscore}. Using these frameworks, we evaluate two metrics, including \emph{similarity} and \emph{correctness}. Similarity measures the semantic similarity between ground truth and $\completion$, while correctness measures answer correctness compared to ground truth as a combination of factuality and semantic similarity. Both metrics range from 0 to 1, with higher values indicating optimal $\completion$. In our case, RAGAs and BertScore is used to evaluate similarity, whereas G-EVAL is used to evaluate correctness of $\completion$. Results of our evaluation is presented in Table \ref{Table:results}.

        %In terms of semantic similarity of $\completion$ with ground truth RAGAs outperformed BertScore for every model. 
        % The best performing model with RAGAs is \emph{prometheus-7b} with highest similarty score of 0.93 and least performing model is \emph{mistral-nemo-minitron-8b} at 0.84. While the best performing model in case of BertScore is \emph{gpt-3.5-turbo} at 0.68 and least being \emph{llama-3.1-8b} at 0.53. In case of correctness of $\completion$ the best performing model was \emph{qwen1.5-7b-chat} at 0.78 closely followed by \emph{gpt-3.5-turbo} and \emph{gpt-4o} at 0.75. While \emph{llama-3.1-8b} reported lowest score of 0.46.
        In our evaluation, the model \emph{Qwen1.5-7B-Chat} performed the best, with the highest similarity score and the lowest standard deviation, as depicted in Fig \ref{fig:evaluation_plot}. On the other hand, \emph{Mistral-NeMo-Minitron-8B-Base} was the least-performing model. We found that `qwen' was the most stable, which is essential in critical domains such as cybersecurity. Contrarily, `mistral-nemo' showed lower accuracy and higher variance. This can be explained through Mistral's sliding attention mechanism that struggles to retail critical information over longer contexts. We also discovered that due to the task criticality, llama 3.1 avoided suggesting a solution, indicating its cautious generation. We observed a similar trend with the `GPT 4o' model. Another critical point to note is that we used a generic instruction prompt for all models, and it is also worth mentioning that model-specific prompt engineering techniques may lead to even better results.

    \subsection{Qualitative Evaluation}
    To justify our quantitative findings (refer to Section \ref{subsec:quantitative_eval}), we qualitatively evaluate the performance of \AlgoName in generating contextually relevant organizational threat intelligence through human evaluation. Given the expensive nature of human evaluation, we engage a panel of \textbf{3} Subject Matter Experts (SMEs), including \textit{one security analyst} and \textit{two cybersecurity researchers}. We task these SMEs to evaluate the \emph{correctness} of generated threat intelligence based on the 58 scenarios and ground truths explained in the preceding section. The SMEs were instructed to rate the correctness of the response on a scale of \textbf{1 to 5}, where 1 represents an incorrect response, and 5 indicates a correct response. We then compare the inter-rater agreement using Fleiss Kappa \cite{mchugh2012interrater} measure. The result of this evaluation shows an agreement score of \textbf{0.6477} with a standard error of \textbf{0.0767}, indicating that the raters' evaluations are not random and are generally aligned, and they \textbf{\emph{substantially agree}} on the \emph{correctness} of the threat intelligence responses generated by \AlgoName. Moreover, \textit{qualitative results} aligns closely with the \textit{quantitative results}, justifying the evaluation.

    % In our context, the mean RAGAS score is reported as \textbf{0.9535} (ranges from 0 to 1, with 1 being the optimal generation), with a standard deviation of \textbf{0.0226} (refer to Figure \ref{fig:evaluation_plot}), highlighting \AlgoName's proficiency in delivering contextually relevant answers within the framework of retrieval-augmented question-answering tasks. 

% ############################################################################################################
\section{\SectionBackground}\label{background} % 1.5 Page

    In the last decade, within the realm of cybersecurity, NLP tasks over unstructured CTI text primarily encompass \textit{Named Entity Recognition, text summarization}, and \textit{analysis of semantic relationships between entities}\cite{rahman2020literature}, etc. Researchers have demonstrated numerous real-world applications using these techniques utilizing CTI gathered from diverse sources \cite{mittal2016cybertwitter,pingle2019relext,piplai2020creating,mitra2021combating}. With the advancement of generative AI in this decade, the application horizon of CTI has proportionally expanded. Liu et al.~\cite{liu2022tricti} introduced a trigger-enhanced CTI (TriCTI) discovery system designed to identify actionable CTI automatically. They utilized a fine-tuned BERT with an intricate design to generate triggers, training the trigger vector based on sentence similarity. Similarly, in \cite{alves2022leveraging}, the researchers employed a BERT classifier to map Tactics, Techniques, and Procedures (TTPs) to the MITRE ATT\&CK framework. On the other hand, Niakanlahiji et al.,~\cite{niakanlahiji2018natural}, proposes an information retrieval system called SECCMiner utilizing various NLP techniques. With SECCMiner, unstructured APT reports can be analyzed, and critical security concepts (e.g., adversarial techniques) can be extracted. A question and answering model called LogQA that answers log-based questions in natural language form using base BERT model and large-scale unstructured log corpora is proposed by Huang et al. ~\cite{huang2023logqa}. Recently, BERT has also been explored to generate contextualized embedding \cite{ranade2021cybert} in cybersecurity. Cybersecurity is a critical domain, and this specialized embedding enables language models to understand the context better. On top of the improvements mentioned, we attempt to integrate LLM to understand the problem context and generate real-time scope-specific threat intelligence while considering different factors. This work is the first attempt to generate complete CTI from diverse sources.

% ############################################################################################################

% ##%%## Add knowledge graph implementation to replace local knowledge retriever
\section{Conclusion}\label{conclusion}
%\sudip{local db consistency issue}
%\sudip{highlight your 3 intro contributions again}
% 0.5 Page
%\manas{At the moment, the conclusion seems similar to Abstract. I would suggest to reshaping it with the following questions: (a) Why would Abridge be a useful tool for SOC? (b) Which component of Abridge helps which functions in SOC? (c) How confident is Abridge compared to its counterpart? (here, a relative percentage improvement would be good) (d) Is the work reproducible? (e) In future work, which component of Abridge would be expanded and how (3-4 lines)?}
%The surge in LLMs has profound implications for cybersecurity, particularly in downstream tasks such as question answering. This paper introduces \AlgoName, a novel framework, %that generates 
%designed to generate %local 
%threat intelligence %(relevant answers) \sudip{why??? is this here in brakets} 
%for a SOC analyst. %, leveraging both global and local cyber knowledge. Comprising \sudip{grammar is wrong}retriever and generator components, the system retrieves global CTI from repositories like CVE, CWE, etc., and local relevant knowledge from local knowledge databases.

%operating % \st{It leverages both global and local cyber intelligence through retriever and generator components. Global knowledge is retrieved from Wikipedia, while local knowledge is extracted from local knowledge databases.} 

This paper introduced \AlgoName, a novel framework that generates contextualized CTI uniquely tailored for an organization depending on its operations. \AlgoName is a valuable tool for SoC analysts due to its unique ability to seamlessly contextualize generic global threat intelligence specific to local operations. The main benefit of this system is its ability to efficiently customize global threat intelligence for local contexts, reducing the need for manual efforts. This gives SoC analysts the necessary information to concentrate on essential tasks, such as developing defensive strategies. We employed qualitative and quantitative evaluations to evaluate \AlgoName's confidence in delivering accurate and relevant threat intelligence. The system exhibited remarkable proficiency in both evaluations, supported by human-generated ground truth responses. It achieved a remarkable RAGAs contextual similarity score of $92\%$ and a correctness score of $78\%$, with a low standard deviation. This underscores the feasibility of automated CTI generation using LLMs and our \AlgoName's robust performance and ability to generate relevant CTI. %consistently. 
In the future, we plan to perform further performance improvement measures, such as developing task-specific retrievers and connecting with cybersecurity knowledge graphs as our local knowledge database for broader evaluations. Additionally, we plan to fine-tune LLMs as part of our performance improvement measures.

%In the future, we aim to enhance the local knowledge by developing a synthetic local knowledge database dataset to improve scalability, and integrate multiple global data sources for enriched global threat intelligence.
%The system leverages both global and local cyber knowledge through retriever and generator components. Global knowledge is retrieved from Wikipedia, while local relevant information is extracted from local knowledge databases. Quantitative evaluation using %the BLEU, ROGUE, and 
%RAGAS %frameworks 
%yielded %mixed 
%impressive results, with %RAGAS outperforming others for answer relevancy with 
%a score of 0.944 for answer relevancy and standard deviation of 0.03534. Future work includes developing a synthetic local knowledge database dataset, incorporating multiple global data sources, and conducting more qualitative evaluations through human assessors.

% ############################################################################################################

\textbf{Acknowledgments.} This work was supported by PATENT Lab at the Department of Computer Science and Engineering, Mississippi State University. The authors would like to thank SME’s for their assistance in qualitative evaluation. The views and conclusions are those of the authors.

\bibliographystyle{splncs04}
\bibliography{sample-base}
\end{document}